# Unraveling a Chemical-Bond-Driven Root of Topology in Three-Dimensional Chiral Crystals

Shungo Aoyagi[1,2,†], Shunsuke Kitou[3], Yuiga Nakamura[4], Motoaki Hirayama[2,5], Hideki Matsuoka[1], Ryotaro Arita[6], Shuichi Murakami[2,7,8], Taka-hisa Arima[3,5] and Naoya Kanazawa[1,†]

[1] *Institute of Industrial Science, University of Tokyo, Tokyo 153-8505, Japan*
[2] *Department of Applied Physics, University of Tokyo, Tokyo 113-8656, Japan*
[3] *Department of Advanced Materials Science, University of Tokyo, Kashiwa, Chiba 277-8561, Japan*
[4] *Japan Synchrotron Radiation Research Institute (JASRI), SPring-8, Hyogo 679-5198, Japan*
[5] *RIKEN Center for Emergent Matter Science, Wako 351-0198, Japan*
[6] *Department of Physics, University of Tokyo, Tokyo 113-8656, Japan*
[7] *Department of Physics, Institute of Science Tokyo, Tokyo 152-8551, Japan*
[8] *International Institute for Sustainability with Knotted Chiral Meta Matter (WPI-SKCM$^2$), Hiroshima University, Hiroshima 739-8526, Japan*

† To whom correspondence should be addressed. E-mail: s-aoyagi@iis.u-tokyo.ac.jp and

  naoya-k@iis.u-tokyo.ac.jp

**Chirality manifests across multiple scales, yielding unique phenomena that break mirror symmetry. In chiral materials, unexpectedly large spin-filtering[1-5] or photogalvanic[6-8] effects have been observed even in materials composed of light elements, implying crucial influence of their topological electronic states[9-13]. However, an underlying framework that links chemical bonding and electronic topology remains elusive, preventing the rational design of quantum chiral properties. Here we identify the chiral bonding network responsible for multifold topological fermions[14,15] by combining synchrotron X-ray diffraction and first-principles calculations on cubic chiral crystals, CoSi and FeSi. Based on the observations of asymmetric valence electron distributions around the transition metals, together with analyses of their bonding to sevenfold-coordinated silicon atoms, we develop a three-dimensional Su-Schrieffer-Heeger model, showing that inter-site hopping on this chiral network creates multifold fermions with doubled topological invariants. Topological features can be switched by reversing the crystalline chirality or tuning electron filling. Our results highlight that implementing strong spin-orbit coupling is not the sole route to realize robust topological phases at elevated temperatures and offer a practical design principle for exploiting chiral topology. Moreover, this real-space framework naturally extends to other elementary excitations or artificial metamaterials, enabling various quantum functionalities through an intuitive approach to chirality engineering.**

Chirality, defined by its distinctive break in mirror symmetry, exerts a profound influence across multiple fields and enables a range of unique functionalities. Left- and right-handed enantiomers, which can arise from identical constituent building blocks, vividly illustrate how two distinct mirror-image structures give rise to asymmetric phenomena, as exemplified by optical activity[16]. On the molecular scale, chiral compounds can distinguish between otherwise nearly identical objects through selective binding or interaction, highlighting the crucial roles of chirality in chemistry and biology[17]. Yet chirality is not confined to the molecular world; it appears in contexts ranging from parity violation in particle physics[18] to spiral shells in biology[19] and even to possible asymmetries in the Universe[20]. The ubiquitous emergence of chirality across vastly different length scales remains an enduring puzzle at the core of fundamental science.

In recent years, focus has intensified on the microscopic quantum behavior of electrons in chiral materials[21,22]. A variety of exotic quantum phenomena have been reported, including chirality-induced spin selectivity (CISS)[1-5], quantized circular photogalvanic effects[6-8], nonreciprocal electrical conduction[23,24] and current-induced magnetization[12,13,25,26], all of which can exhibit unusually large responses even in materials composed of relatively light elements. These findings hint that robust chiral features may emerge without requiring strong spin-orbit coupling.

Recent theoretical and experimental studies suggest that the unexpectedly large quantum responses in chiral systems may be driven by the presence of topological band degeneracies at time-reversal-invariant momenta (TRIM)[9,10,27-30]. Such degeneracies generate giant Berry curvature and orbital magnetic moments[11-13,31], potentially boosting such chiral quantum responses. With the development of group-theoretic classification and comprehensive material catalogs[32-35], the topological aspects of chiral crystals have been also systematically mapped. However, the rational design of giant quantum responses remains limited by a lack of real-space chemical bonding insights. This gap hinders intuitive strategies for material development and discoveries of new chiral systems.

To bridge this gap, we used synchrotron X-ray single-crystal diffraction to directly visualize valence electron density (VED) distributions in cubic chiral crystals, providing non-empirical evidence of their asymmetric charge spread. Our first-principles calculations revealed that structural chirality creates an alternating network of bonding and antibonding states, which underlies the observed asymmetric VED distribution. Integrating these findings, we developed a three-dimensional (3D) version of the Su-Schrieffer-Heeger (SSH) model to demonstrate how the band structure acquires a chiral twist in its geometric phase. This reframes the known doubling of the topological invariant in chiral systems[14,15] from a chemical bonding-based perspective and enables the rational design of chiral topology.

**Visualization of valence electron density**

To experimentally probe the chemical bonding states in chiral crystals, we examined the VED of CoSi and FeSi single crystals using high-photon-energy X-ray diffraction in combination with the recently developed core differential Fourier synthesis (CDFS) method[36,37] (Methods and Supplementary Note 1). These transition metal silicides are members of the $B$20-type compounds, characterized by a cubic chiral structure (space group $P2_13$). We comprehensively measured the diffraction intensities of a homochiral crystal over a wide range of reciprocal space (Fig. 1**a**; also see Supplementary Note 1). We then subtracted the core electron density contributions from the observed intensities using the CDFS method[36,37], which enabled us to isolate and directly visualize the VED primarily involved in chemical bonds. Notably, CDFS provides a powerful real-space approach to directly evaluate the electronic structure, offering chemical bonding insights that complement the established momentum-space observations.

Figure 1**b** shows observed VED iso-density surfaces around one of the four symmetry-equivalent cobalt sites in the cubic unit cell of CoSi (predominantly left-handed; see Supplementary Note 1). In the outer regions at lower electron densities (2.5 e/Å$^3$ and 6.0 e/Å$^3$), the isosurfaces are nearly spherical and isotropic. In contrast, at a higher density of 11.0 e/Å$^3$, the inner isosurface becomes markedly anisotropic, manifesting the asymmetric nature of the VED distribution. In particular, contrary to naive expectations, a larger portion of the VED resides on the opposite side to the nearest-neighbor (NN) Si atom situated on the ⟨111⟩ threefold rotation axis. By analyzing the VED in several cross-sections perpendicular to the [111] axis (horizontal lines in Fig. 1**b**), we revealed that the distribution maintains threefold rotational symmetry, while twisting clockwise as it extends along $[\bar{1}\bar{1}\bar{1}]$ (Fig. 1**c**). Measurements on a crystal with the opposite chirality (predominantly right-handed) showed a mirror-image electron density across a plane parallel to [111] (Fig. 1**d**), exhibiting the reversed sense of twisting. These results demonstrate that the crystalline handedness is directly imprinted in the VED.

Next, we investigate the impact of electron filling levels on the asymmetric VED by comparing left-handed crystals of CoSi and FeSi. Replacement of the transition metal element in the *B*20 silicide family leaves the band structure intact, while shifting the Fermi level in accordance with the number of valence electrons[38]. This validity of rigid band approximation ensures FeSi as a suitable reference system for this purpose. We consider the sevenfold coordination of silicon atoms around the transition metal site (Fig. 2**a**) to highlight the role of chemical bonding in the spatial imbalance of VED (Fig. 2**b** for CoSi; Fig. 2**c** for FeSi). While the VED (the isosurface of 8.43 e/Å$^3$) is asymmetric in FeSi, keeping the threefold symmetry as well, its skewness differs markedly from that of CoSi. Specifically, the high-density region is reversed to the side of NN Si site in FeSi. [Compare the VEDs of CoSi and FeSi from the [111] direction (left view) in Figs. 2**b**-1 and 2**c**-1.] Cross-sectional VED visualizations highlight small yet distinct charge imbalances for each bond direction (Figs. 2**d** and 2**e**). In FeSi, the VED is larger toward the second-nearest-neighbor (SNN) Si site than the third-nearest-neighbor (TNN) site (Fig. 2**d**). In contrast, in CoSi, the VED appears disproportionately high toward the TNN site (Fig. 2**e**). The TNN bond length is correspondingly shorter in CoSi (2.45 Å) than in FeSi (2.52 Å), implying that one extra electron in CoSi is mainly accommodated in bonding states along the TNN directions. Altogether, the bonding strength and orientation are strongly modified according to the filling levels of *d*-orbitals extending to different Si neighbors, which results in the large variation in the asymmetric VED with the Fermi level shifts.

**Asymmetric bonding states**

To support our experimental findings, we quantified the metal–silicon bonding strength based on the first-principles electronic structure calculations (see Methods). Figures 3**a** and 3**b** show the band structure and total density of states (DOS) for CoSi and FeSi. Even without spin-orbit coupling, the band structure hosts multifold point degeneracies, including threefold point degeneracy at Γ and double twofold degeneracies at R (Fig. 3**a**)[14,15]. Notably, in CoSi, these degeneracies are located near the Fermi energy $E_F$, offering an ideal platform for exploiting the topological properties with minimal influence of the trivial hole pocket at M. To separate the contributions of different $d$-orbitals, we decomposed the total DOS into partial DOS for three groups: $d_{3z^2-r^2}$, $d_{xy} + d_{x^2-y^2}$, and $d_{xz} + d_{yz}$, considering the threefold symmetry about the [111] axis (Fig. 3**c**). Here, the [111] direction is defined as the $z$-direction. The partial DOS of $d_{3z^2-r^2}$ remains nearly zero above $E_F$ of FeSi, representing that this orbital is fully occupied. It follows that the topological band structure around $E_F$ of CoSi is not comprised of the $d_{3z^2-r^2}$ orbital, which primarily forms covalent bonds with the NN Si located along the [111] axis.

Next, we focus on the Fe $d$-orbital bonding states with SNN and TNN Si atoms by calculating the crystal orbital Hamilton population (COHP)[39-41], as shown in Fig. 3**d**. The COHP evaluates the Hamiltonian matrix elements for the orbitals involved in the targeted chemical bonds, providing a quantitative measure of bond strength and its sign. By convention, the COHP is expressed with the minus sign so that bonding (antibonding) contributions are positive (negative). As the filling level increases across $E_F$ of FeSi, the character of the SNN bonds changes from bonding to antibonding, whereas the TNN bonds exhibit the opposite trend, shifting from antibonding to bonding. Given the complete occupation in $d_{3z^2-r^2}$, the reversal in the skewed distribution between CoSi and FeSi (Fig. 2) is attributed to changes in the bonding characters of SNN and TNN bonds. Specifically, more VED accumulates on the same (opposite) side to the NN Si site when the SNN and TNN bonds are bonding and antibonding (antibonding and bonding) in FeSi (CoSi), as schematically illustrated in Fig. 3**e**. This multi-orbital analysis therefore pinpoints two inequivalent transition-metal–Si hoppings as the decisive drivers of the observed VED asymmetry.

## 3D SSH model for chiral topology

Based on the chemical bonding states identified above, we develop a 3D SSH model that isolates the essential bonding network formed by the SNN and TNN bonds, thereby illustrating the emergent topological band structure. Figure 4**a** shows the extracted bonding network after omitting the NN bonds. Along the [001] direction (upward in Fig. 4**a**), the SNN and TNN bonds (*i.e.*, bonding and antibonding states, or *vice versa*) alternate to form 1D chains. This arrangement is reminiscent of the SSH model[42], which describes the topological electronic state in a polyacetylene with alternating single and double bonds. By considering the cubic symmetry, we extend this concept to a fully 3D SSH lattice composed of spinless single orbitals (Fig. 4**b**). The model includes only two types of hopping coefficients ($v$ for SNN and $w$ for TNN), and on-site potentials ($+\mu$ for transition metal and $-\mu$ for silicon). In particular, when viewed along the [111] axis, this 3D bonding network hosts a windmill-like arrangement that reproduces the chiral symmetry ($P2_13$) of the native atomic lattice (Fig. 4**c**).

This drastic reduction from a multi-orbital framework to a single-orbital 3D SSH lattice constitutes a key conceptual advance, revealing that the chiral topological bands arise from the asymmetric pattern of bonds. The calculated band structure in this model ($v = 0.1$, $w = 0.3$, $\mu = 0.1$) replicate the first-principles results (compare Fig. 4**d** with Fig. 3**a**), including the point degeneracies at TRIMs. Investigation of the band topology further confirms that the threefold bands at Γ carry topological charges of $C = \pm 2$ and 0, while the double twofold bands at R have $C = \pm 1$ in pairs, consistent with the conventional symmetry-based arguments (Supplementary Note 2).

Scrutinizing the quantum phase structure, we gain deeper insight into the intrinsic connection between chirality and topology. We focus on the lowest branch of the conduction band (the upper green branch in Fig. 4**d**) and evaluate its Berry phase along an infinitesimally small loop $\lambda_1$ (red arrow in Fig. 4**e**) that connects four points related by fourfold rotation around the Γ-R line. With a specific gauge choice, this loop yields a Berry phase of $-2\pi$ (Fig. 4**f** and Supplementary Note 2). A real-space depiction of the phase structure shows that the atomic orbitals, which are located at sites linked by the threefold rotation, exhibit distinct phase values $\theta_i$, ($i$ = 1, 2, 3) (Fig. 4**f** and Supplementary Note 2). Namely, a chiral arrangement of phases emerges as a consequence of the absence of mirror planes. Under successive threefold rotations along the loop, these phases rotate in sequence to produce a total geometric phase of $-2\pi$ (Fig. 4**f**). When tracing the inverted loop $\lambda_2$ with respect to Γ (blue arrow in Fig. 4**e**), the chiral phase patterns undergo the sign reversal, again accumulating a $-2\pi$ geometric phase (Fig. 4**g**). Given that the wavefunction are smoothly connected within the cylindrical region connecting $\lambda_1$ and $\lambda_2$

(Fig. 4**e** and Supplementary Note 2), two line-singularities of the gauge potential emanate along the $\Gamma$-R and $\Gamma$-$\bar{\text{R}}$ lines, thereby doubling the Chern number.

In a similar manner, the other conduction-band branches exhibit oppositely-handed chiral or achiral phase structures, yielding the opposite topological charge ($C = -2$) or no phase winding ($C = 0$), respectively. Here we emphasize that the emergence of topological fermions is rooted solely in the inter-site hopping on the 3D chiral network of covalent bonds. This essence also highlights the irrelevance of spin-orbit couplings, providing a simple chemical design principle for achieving robust topological phases at elevated temperatures without relying on heavy elements.

Finally, we construct a phase diagram for the topology of the multifold fermions at $\Gamma$ (Fig. 4**h**). Changing the hopping ratio $v/w$ across the critical value $v/w = 1$ (shifting along the horizontal axis in Fig. 4**h**) triggers a sign reversal of the topological charge. Since crossing across $v/w = 1$ can be regarded as interchanging the SNN and TNN bonds, the observed transition represents that the band topology can be switched by the reversal of crystalline chirality. In the $v/w = 1$ state, which corresponds to an achiral structure similar to the NaCl-type lattice, the electronic structure becomes topologically trivial.

Electron filling (the position of $E_F$ in Fig. 4**h**) additionally alters the topology. As already indicated in the band structure (Fig. 4**d**), the sign of the topological charge is opposite between valence and conduction bands due to the particle-hole symmetry (Supplementary Note 2). This behavior also aligns with the switch between bonding and antibonding characters on the SNN and TNN bonds across the Fermi level of FeSi (Fig. 3**d**). In other words, while the *crystalline chirality* remains fixed, the *electronic chirality* can be reversed simply by shifting the Fermi level.

**Conclusion and Outlook**

Our work demonstrates that the crystalline chirality is directly transcribed into the chemical bonding network, as evidenced by the twisted VED distributions. The topological properties emerge because the wavefunctions consequently acquire a handedness in their phase structures through asymmetric inter-site hopping, leading to pronounced Berry curvatures and orbital magnetic moments. We further show that the topological features can be controllably switched not only by structural chirality but also by tuning the bonding or antibonding characters via electronic filling.

A key development is the application of CDFS to visualize real-space VED distributions, complementing conventional momentum-space observations of topological states. By offering direct insights into chemical bonding, CDFS can serve as a powerful tool for rational materials design, especially where a real-space perspective is indispensable. Indeed, our 3D SSH model founded on CDFS identifies the chemical-bond-driven root of the chiral topology, suggesting a path to circumvent the reliance on strong spin-orbit interactions and enabling robust quantum properties in lighter-element chiral systems.

Our theoretical framework can expand in several promising directions. On the fundamental side, the inclusion of strong electron correlations, magnetic orders and structural defects offers exciting possibilities to explore novel quantum phases where topology and interaction effects intersect. Such extensions possibly guide the design of materials with tailor-made properties such as exotic superconductivity[43] and chiral magnetism[44]. Beyond the realm of electron-based systems, the concept of a chiral topological network extends naturally to various elementary excitations[45,46], artificial metamaterials[47-49], and quantum simulations[50], opening up a wide range of opportunities to control wave propagation and topological functionalities through chiral engineering.

## Methods
### Sample preparation
The single crystals of CoSi and FeSi were grown by the chemical vapor transport (CVT) method[51]. Stoichiometric amounts of the transition metal and silicon elements were arc-melted to achieve a homogeneous mixture. The resulting melt was then ground into a powder form, which was loaded into a quartz tube along with a transport agent ($I_2$) and subsequently sealed under vacuum. For CoSi, a temperature gradient was applied with the high-temperature zone at 1000 °C and the low-temperature zone at 900 °C, whereas for FeSi, the high- and low-temperature zones were set at 900 °C and 800 °C, respectively. The samples were grown under these conditions for three weeks, typically yielding single crystals on the order of 100 μm in size. The crystalline chirality is defined such that, in the left-handed (right-handed) crystal, the transition metal atoms are stacked in a clockwise (counterclockwise) arrangement along the [111] direction. The crystal structures are drawn by using VESTA[52].

### XRD experiments
The XRD experiments were performed on BL02B1 at a synchrotron facility SPring-8, Japan[53]. An $N_2$-gas-blowing device was used to cool the crystal to 100 K. A two-dimensional detector CdTe PILATUS, which had a dynamic range of ~$10^6$, was used to record the diffraction patterns. The intensities of Bragg reflections of the interplane distance $d > 0.28$ Å were collected by the CrysAlisPro program[54] using a fine slice method, in which the data were obtained by dividing the reciprocal lattice space region in increments of ω = 0.01°. Intensities of equivalent reflections were averaged, and the structural parameters were refined by using JANA2006[55]. High-angle reflections (sin $\theta/\lambda > 0.5$ Å$^{-1}$) were exclusively used for structural refinement to perform high-angle analysis. Since the contribution of spatially spread valence electrons is very small in the high-angle region, the structural parameters, including the atomic displacement parameters (mainly due to the thermal vibration), are obtained with high accuracy.

**CDFS analysis**

The CDFS method was used to extract the VED distribution around each atomic site[36]. Core electron configurations were assumed as [Ne] for Si, and [Ar] for both Fe and Co atoms. The contribution of the thermal vibration was subtracted from the VED using the atomic displacement parameters determined by the high-angle analysis[36]. In CoSi, the voxel of the three-dimensional electron distribution is defined as a cube with a side length of 0.0222 Å, and in FeSi, it is defined as a cube with a side length of 0.0224 Å. It should be noted that the absolute value of the obtained electron distribution does not directly reproduce the number of valence electrons around the atoms, partly because the double scattering, absorption, extinction, and detector saturation[56] could not be completely excluded in the measurement of diffraction intensities. Although the analysis may suffer from the indefinite phase problem for weak reflections, the same anisotropy in the VED distribution was consistently observed across multiple samples. Furthermore, in crystals of opposite chirality, the chirality of the VED distribution was also reversed.

**First-principles calculations**

The first-principles calculations were performed in the Vienna Ab initio Simulation Package (VASP) without the spin-orbit coupling[57,58]. The pseudopotentials were generated with the projector augmented wave method, and the Perdew-Burke-Ernzerhof (PBE) type within the generalized gradient approximation (GGA) framework was applied for the exchange-correlation functions. The crystal structure used for the calculations was obtained from the high-angle XRD data. A 12×12×12 k-point mesh and a 500 eV plane-wave energy cutoff were adopted. The electronic self-consistent field (SCF) calculations were converged until the total energy changed by less than 1×10$^{-6}$ eV. The calculation of the band structure was performed with Vaspkit[59]. The calculations of density of states (DOS), partial density of states (pDOS), and crystal orbital Hamilton population (COHP) method were performed with the LOBSTER software package[60].

**Acknowledgements** We thank M. Hirschberger and R. Misawa for experimental supports. This work was supported by JSPS KAKENHI (Grants No. 21H04990, No. 22H00108, No. 23H04017, No. 23H05431, No. 23H05462, No. 24H00417, No. 24H01644, No. 24H02231, No. 24H01652, No. 25H01246, No25H01252 and No. 25H02126), JST FOREST (Grants No. JPMJFR2038, No. JPMJFR2362), JST CREST (Grants No. JPMJCR23O3 and No. JPMJCR23O4), the Mitsubishi Foundation, the Sumitomo Foundation, and FoPM (WINGS Program, the University of Tokyo). The synchrotron radiation experiments were performed at SPring-8 with the approval of the Japan Synchrotron Radiation Research Institute (JASRI) (Proposals No. 2024B2010 and No. 2024B2017)

**Author contributions** S.A. and N.K. grew the single crystals. S.A., S.K. performed XRD measurements with support from Y.N., T.A. and N.K.; S.A., M.H. and R.A. performed first-principles calculations. S.A. developed the SSH model with support from H.M., S.M. and N.K.; S.A. and S.M. performed the analysis of electronic topology. N.K. organized the project. S.A. and N.K. wrote the draft and all the authors discussed the results and commented on the manuscript.

**Competing interests** The authors declare no competing interests.

**Figures**

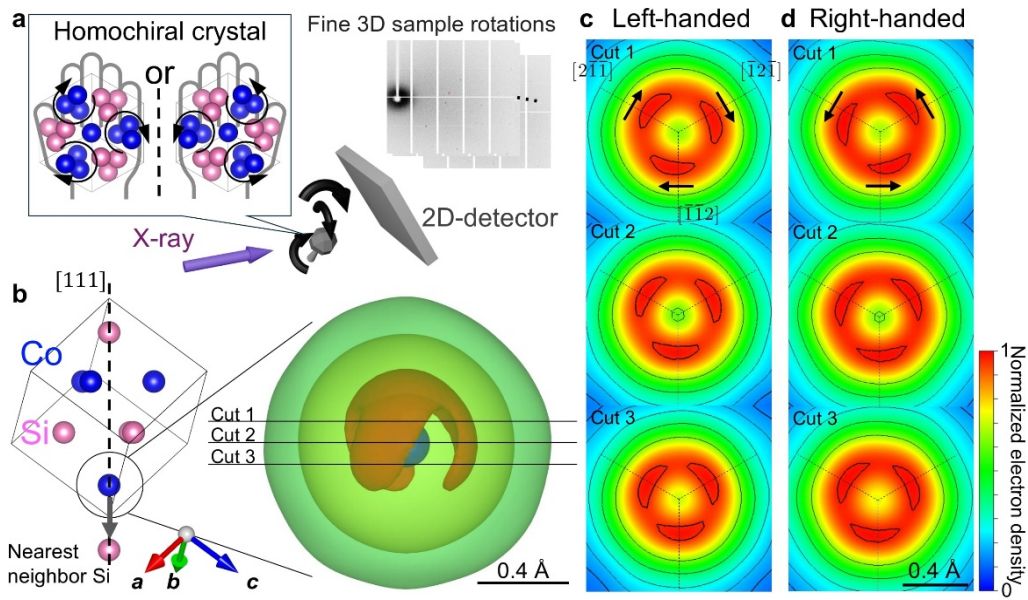

**Fig. 1 | Experimental setup and chiral VED observed in CoSi.**

**a,** Schematic of the synchrotron X-ray diffraction experiment on a homochiral single crystal. A total of 54,000 diffraction patterns were collected under crystal rotation to enable valence electron density (VED) visualization via core differential Fourier synthesis (CDFS). **b,** Unit cell of left-handed CoSi (left) and the visualized VED around a Co-site (right). Green, yellow and red isosurfaces represent electron densities at 2.5, 6.0 and 11.0 $e/Å^3$, respectively. The blue sphere at the center of VED represents the position of the Co nucleus. The scale bar represents 0.4 Å. **c, d,** Cross-sectional views of the VEDs for the left- (panel **c**) and right-handed (panel **d**) crystals. The horizontal lines in panel **b** mark the positions of the cross-sections. The scale bar represents 0.4 Å.

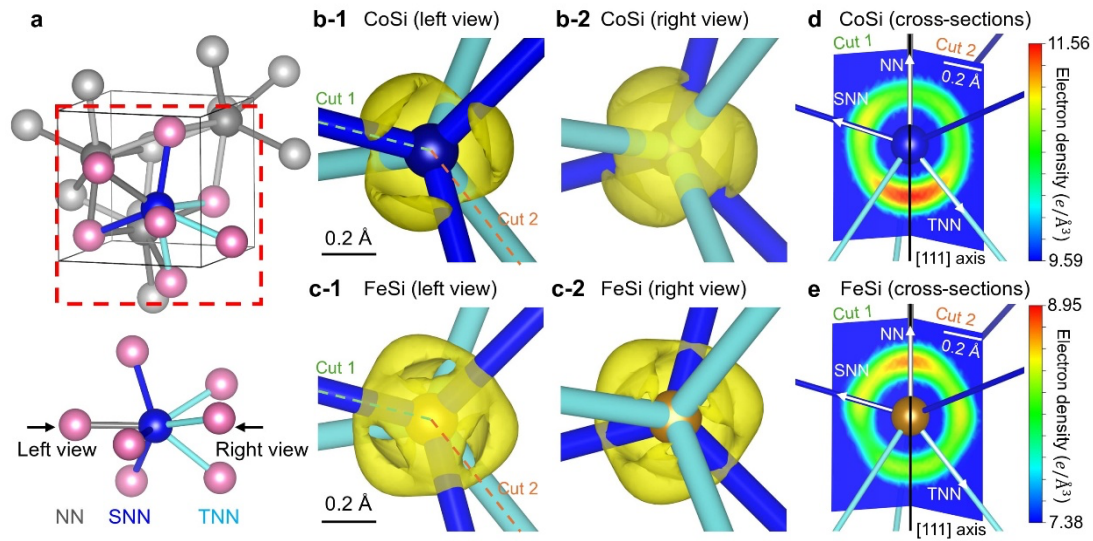

**Fig. 2 | Variation in asymmetric VED with electron filling.**

**a,** Coordinated structure around a transition metal site in a B20-type crystal. The transition metal atom is bonded to seven neighboring silicon atoms: one nearest neighbor (NN, gray bond; 2.31 Å in CoSi and 2.29 Å in FeSi), three second nearest neighbors (SNN, blue bonds; 2.34 Å in CoSi and 2.35 Å in FeSi), and three third nearest neighbors (TNN, sky blue bonds; 2.45 Å in CoSi and 2.52 Å in FeSi). **b, c,** Isosurfaces of the VEDs around the Co site in CoSi at 11.00 $e/Å^3$ (**b-1**, **b-2**: left and right views) and the Fe site in FeSi at 8.43 $e/Å^3$ (**c-1**, **c-2**: left and right views). Viewing directions are indicated in panel **a**. **d, e,** Cross-sectional VED images on planes that include the NN-SNN bonds (left side; cut 1) and the NN-TNN bonds (right side; cut 2). The scale bar represents 0.2 Å. The dashed lines in panels **b-1** and **c-1** mark the positions of the cross-sections. Note that the VED color scales differ between **d** and **e**, and CoSi displays overall higher VED values than FeSi.

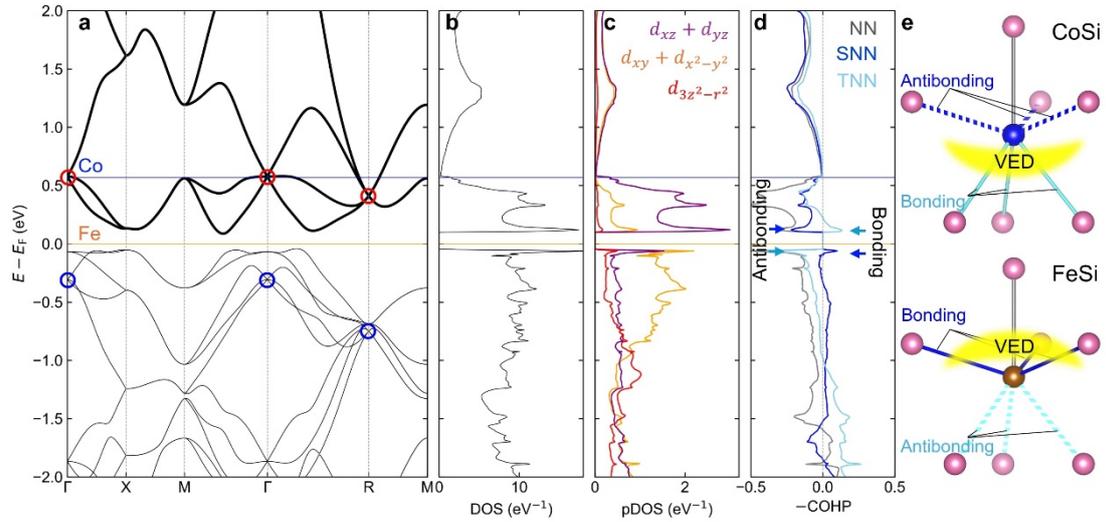

**Fig. 3 | Bonding-state analyses based on band-structure calculations.**
**a, b,** Band structure and total density of states (DOS) of FeSi without spin-orbit coupling. The Fermi level for CoSi is also shown, based on the rigid band approximation. Bold-lined band branches and circled topological point degeneracies highlight the features reproduced by the single-orbital 3D-SSH model and revisited in Fig. 4d. **c,** Partial DOS of three groups of $d$-orbitals: $d_{3z^2-r^2}$ (red), $d_{xy} + d_{x^2-y^2}$ (orange) and $d_{xz} + d_{yz}$ (purple). The $z$-direction corresponds to the threefold axis at the transition metal site, *i.e.*, [111] direction. **d,** Crystal orbital Hamilton population (COHP) analysis for three bonding directions: NN (gray), SNN (blue) and TNN (sky blue). The horizontal axis plots $-$COHP, so that bonding (antibonding) states appear at positive (negative) values. **e,** Schematic illustrations of the bonding characters between the transition metal and silicon atoms in CoSi (upper) and FeSi (lower). The NN bond is nearly fully occupied, whereas the SNN and TNN bonds exhibit bonding (solid lines) or antibonding (dashed lines) states depending on the electron filling. A greater amount of VED (yellow shades) accumulates in the bonding region. (Also see Figs. 2d and 2e.)

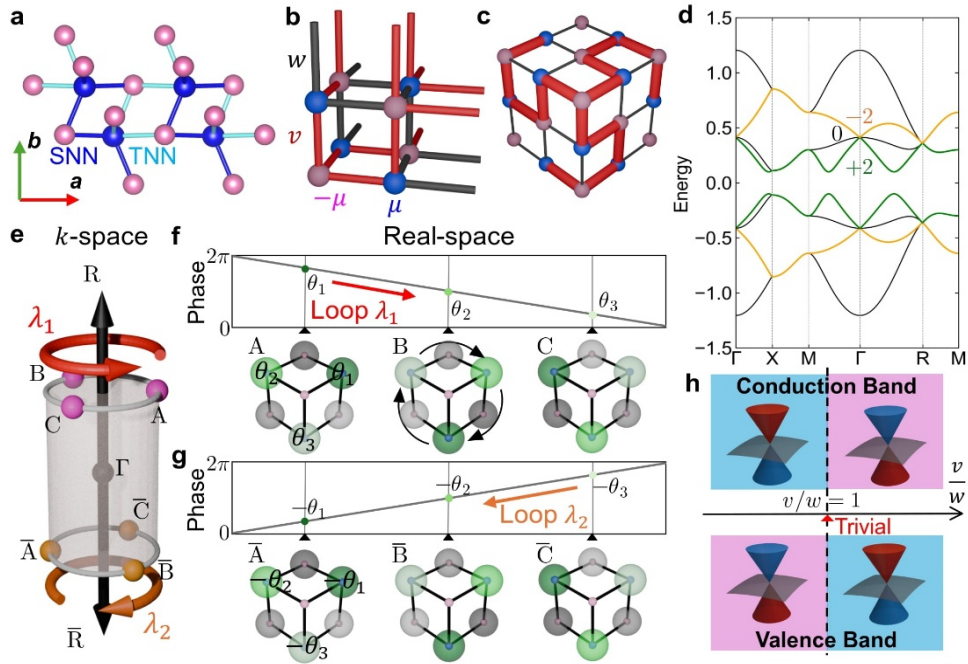

**Fig. 4 | Real-space depiction of multifold topological fermions and tunable chiral topology.**

**a,** Extracted bonding network comprising SNN and TNN bonds in CoSi, which governs the electronic structure near the Fermi energy. SNN bonds (solid blue lines) bear antibonding character, whereas the TNN bonds (solid sky-blue lines) exhibit bonding character, forming an alternating bonding-antibonding pattern reminiscent of the SSH model. **b,** Implementation of the alternating pattern into a 3D SSH model on a simple cubic lattice, with hopping parameters $v$ and $w$, and on-site potentials $\pm\mu$. **c,** Chiral bonding network of the 3D SSH model viewed along the threefold axis. The thick red lines denote inter-site hopping characterized by the parameter $v$. **d,** Band structure of the 3D SSH model, highlighting the threefold band branches at Γ with Chern numbers $C = +2$ (green), $-2$ (orange). **e,** Two small loops $\lambda_1$ and $\lambda_2$ in momentum space for Berry phase analysis. **f, g,** Evolution of Berry phases and the real-space phase configurations of wavefunctions along $\lambda_1$ (**f**) and $\lambda_2$ (**g**). The chiral configuration of phases with three distinct values (spheres dyed with greenish or grayish tones) rotate clockwise (counterclockwise) when tracing the closed loop $\lambda_1$ ($\lambda_2$), yielding a $2\pi$ Berry phase. **h,** Phase diagram of the threefold topological fermions at Γ with respect to crystalline chirality ($v/w$) and electron filling (*i.e.*, whether the Fermi level $E_F$ lies in the valance band or the conduction band). The discrete transition in topological invariant can be induced by varying either $v/w$ or $E_F$. A trivial phase appears only at $v/w = 1$ (achiral state).